\documentclass[aps,preprint,secnumarabic,nobalancelastpage,nofootinbibt]{revtex4}

\usepackage{graphics}      
\usepackage{graphicx}      
\usepackage{longtable}     
\usepackage{url}           
\usepackage{bm,color}            
\usepackage{amssymb, amsmath, enumerate, theorem, epsfig,subfigure,setspace}

\newcommand{\nb}[1]{\textcolor{black}{#1}}

\begin{document}

\title{Polaritonic network  as a paradigm for dynamics of coupled oscillators}

\author{Kirill P. Kalinin${}^1$ and Natalia G. Berloff${}^{2,1}$ }
\email[correspondence address: ]{N.G.Berloff@damtp.cam.ac.uk}
\affiliation{${}^1$ Department of Applied Mathematics and Theoretical Physics, University of Cambridge, Cambridge CB3 0WA, United Kingdom }
\affiliation{${}^2$ Skolkovo Institute of Science and Technology,  Bolshoy Boulevard 30, bld. 1 Moscow,121205 Russian Federation}

\date{\today}

\begin{abstract}{ Photonic and polaritonic lattices have been recently theoretically proposed and experimentally realised as many-body simulators due to the rich behaviors exhibited by such systems at the macroscale. We show that the networks of polariton condensates  encapsulate a large variety of behaviours of systems of coupled oscillators. By eliminating spatial degrees of freedom in nonresonantly pumped polariton network, we establish that depending on the  values of experimentally tunable parameters the networks of polariton condensates may represent Kuramoto, Sakaguchi-Kuramoto, Stuart-Landau, Lang-Kobayashi oscillators and beyond.   The networks of polariton condensates are therefore capable of implementing various regimes acting as analogue spin Hamiltonian minimizers, producing complete and cluster synchronization, exotic spin glasses and large scale secondary synchronization of oscillations. We suggest that the recently implemented control of the system parameters for individual sites in polariton lattices allows addressing for the first time the interaction of sublattices  that belong to different oscillatory classes. }
\end{abstract}

\maketitle

For a long time, two pervasive topics of modern science -- dynamics of coupled oscillators and simulations of many-body solid state systems  -- have barely crossed each other's paths.
Complex dynamic behaviour of networks of coupled oscillators arises  in various scientific disciplines ranging from biology, physics, and chemistry to social and neural networks as well as in established and emerging technological applications. Such networks  served as paradigmatic models for understanding 
the mechanism of various  collective phenomena. For instance, the Kuramoto oscillators \cite{kuramoto,kuramoto2}  have been successfully used to represent,  study or even predict a wide variety of pattern \nb{formation}  in spatiotemporal systems, such as biochemical systems, neural networks, convecting fluids, and laser arrays. The reason for such power of networks of coupled oscillators in describing vastly different systems lies in the underlying symmetries of the system:  these are described by similar universal order parameter equations  that share similar characteristics \cite{halperin}.
Such symmetries make it possible to divide systems into various universality classes that differ only by  the nature of the dynamics \cite{kadanoff} and allow  one not only to draw similarities between very different physical systems but also  {\it  predict} the behaviour of the new systems that fall into previously known universality class \cite{KBuniversal}. 
 
Traditionally, at the other end of the spectrum of nonlinear dynamical studies lie the complex many-body solid-state systems that are often considered as powerful platforms for simulating  various elaborate Hamiltonians. A number of systems were realised implementing lattices of various physical origin:
 neutral atoms, ions, electrons in semiconductors, polar molecules, superconducting circuits, nuclear spins etc. \cite{QSreview}. These are typically equilibrium systems that realise  ground or excited states of their structure  Hamiltonians. Recently, photonic and polaritonic lattices have emerged as promising platforms for many-body quantum and classical simulations \cite{photonlattices2007,bloch2008}. These systems are typically of a gain-dissipative nature, capable of symmetry breaking and spontaneous pattern forming, and have constant nonzero particle fluxes even at the steady state.  Furthermore, as we argue in this article when the lattice elements  have photonic component and the gain-dissipative nature the  wavefunction packets evolve, interact and synchronise in a close resemblance to the coupled oscillators that are governed by the universal  order parameter equations. As a result, on the one hand, many classical phenomena found in such lattices can be explained or predicted by the behaviour of the corresponding system of coupled oscillator networks from the same universality class; on the other hand, strong nonlinearities,  spin polarization, sensitivity to magnetic fields, and individual site control greatly enrich possible states and dynamical regimes that can be generated in such lattices. 
In this article, we propose and theoretically justify the use of networks of exciton-polaritons ({\it polaritonic networks}) as a flexible universal platform to realise a vast array of known and extensively studied systems of coupled oscillators, to probe new exotic dynamical regimes and to create novel states of matter \nb{that may result from hybridisation of several different networks on one platform}. In particular, we find the states that suggest  possible practical implementations towards  optical transistors  and clarify the requirements for such a network to minimise  classical  spin Hamiltonians.

In the last decade, it emerged that strong light-matter interactions in semiconductor  microcavities offer a versatile platform to realise nontrivial states, dynamics, localised structures. They consist of exciton-polaritons (polaritons) that  are bosonic quasi-particles  with a tiny effective mass which is typical $10^{-4}-10^{-5}$ of the bare exciton mass. Their  energy-momentum dispersion curves can be controlled by appropriate detuning, and their properties and dynamics can be readily accessed by angular-resolved photo- or
electroluminescence  spectroscopy. A  wealth of experimental results have been demonstrated with these systems, including  Bose-Einstein condensates (BEC) \cite{bec},  polariton lasers \cite{lasers}, polariton parametric amplifiers \cite{ppa}, and cavity quantum electrodynamics \cite{cavity}. The polariton BEC or lasing has been demonstrated in various materials such as CdTe \cite{bec}, GaAs \cite{gaas,balili2007}, GaN \cite{gan}, organic polymers \cite{organic} and using  optical pumping or  electrically pumped exciton-polariton emitters \cite{yamamotoElPump}.   

We are interested in networks of $N$ polariton condensates  created at  lattice sites  -- the vertices of a two-dimensional graph at positions ${\bf r}_i, i=1,\cdot\cdot\cdot, N.$ Many techniques are available to  engineer a variety of the potential  landscapes of polaritons \cite{chneiderRev, amo}. Polariton can be confined by strain-induced traps \cite{snoke}, surface acoustic waves \cite{Cerda_PRB2012}, direct fabrication with the gold deposition technique \cite{Kim_NatPhys2011}, by using hybrid air gap microcavities \cite{Daffer}, or by coupled mesas etched during the growth of the microcavity \cite{Winkler_NewJPhys2015}, by micropillars  \cite{micropillars} and  in various geometries: square \cite{Kim_NatPhys2011}, triangular \cite{Kim_NewJPhys2013}, hexagonal \cite{Kusudo_PRB2013}, fully etched honeycomb \cite{Jacqmin_PRL2014}, Kagome \cite{Masumoto_NewJPhys2012} or even in quasi-periodic potentials \cite{amoFibb}.    However, the potential traps in a gain-dissipative system lead to complicated dynamics as the flow dynamics  present in  gain-dissipative systems even in the steady state is highly nontrivial in this geometry. To avoid such complications, polarion lattices can be optically engineered by exploiting the  interactions between polaritons and reservoir excitons that can be injected  in specific areas of the sample. Excitons barely move from the point where they are excited as they are orders of magnitude heavier than polaritons. Experimentally, the lattice is achieved by  using a spatial light modulator that creates polariton condensates at the vertices of any prescribed graph \cite{wertz,manni,pendulum,baumbergGeometrical,BerloffNatMat2017}. This technique also allows  controlling the intensities, $p_i,$  of individual sites indexed by $i$ depending on the density of the polariton condensate at this site, if needed. In what follows we derive and discuss the network behaviour bearing in mind this technique of the lattice formation, however, other ways to create polariton lattices affect the parameters, but not the universality of the derived coupled oscillators equations.

The mean-field behaviour of polariton condensates is  described  by the generalised  complex Ginzburg-Landau equation  (cGLE)  (often also referred to as a driven-dissipative Gross-Pitaevskii equation) coupled to the reservoir dynamics \cite{goveq, carusotto, reviewCarusotto}. Although the process of Bose-Einstein condensation includes quantum effects, when condensate is formed it is  accurately described by the mean-field equations as was shown in numerous experimental works \cite{meanfield,meanfield2,meanfield3,meanfield4,meanfield5,meanfield6,meanfield7,meanfield8,meanfield9}.  The   equation on the wavefunction $\psi({\bf r},t)$ of the condensed system is coupled to the rate equation on the density of the hot reservoir $n_R({\bf r},t)$ so that 
$	
	i  \psi_t = - \frac{1}{2 m}(1 - i \hat{\eta} n_R)  \nabla^2\psi + U_0 |\psi|^2 \psi+
g_R n_R \psi 	  +\frac{i}{2} [R_R n_R - \gamma_C] \psi,$ and 
	  $n_{Rt}=  - \left(\gamma_R+ R_R|\psi|^2 \right) n_R + P({\bf r},t),$
	where we set $\hbar=1$, $U_0$ and $g_R$ are the polariton-polariton and polariton-exciton  interaction strengths respectively, $\hat{\eta}$ is the energy relaxation \cite{wouters12,KBuniversal}, $R_R$ is the rate of scattering from the hot reservoir into the condensates. The condensate ($\gamma_C$) and the reservoir ($\gamma_R$) relaxation rates describe photon losses in the cavity and hot exciton losses other than scattering into condensates.  The incoherent pump source is described by  the pumping intensity $P({\bf r},t)$. We nondimensionalise these equations  by $\psi \rightarrow \sqrt{ \gamma_C / 2 U_0} \psi$, $t \rightarrow 2 t / \gamma_C$, $ {\bf r} \rightarrow \sqrt{1 / m \gamma_C} {\bf r}$, $n_R \rightarrow \gamma_C n_R / R_R, P \rightarrow P\gamma_C^2/2R_R $ and introduce the dimensionless parameters  $g= 2 g_R / R_R$, $b_0= 2 \gamma_R / \gamma_C$, $b_1=  R_R / U_0$,  $\eta=\hat{\eta} \gamma_C / R_R $. The resulting model yields
 \begin{eqnarray}	
	i  \frac{\partial \psi}{\partial t} &=& - (1 - i \eta n_R)  \nabla^2\psi +  |\psi|^2 \psi+
g n_R \psi 	  + i(n_R - 1) \psi, \label{e1}\\
	  \frac{\partial n_R}{\partial t} &=&  - \left(b_0+ b_1|\psi|^2 \right) n_R + P({\bf r},t).
	\label{e2}
\end{eqnarray}
A  unique property of the exciton-polariton  system  is the flexibility with which the parameters $g, b_0, b_1, \eta$ can be controlled and changed to allow the system span various regimes bridging   lasers or other nonequilibrium systems with equilibrium condensates and entering novel physical regimes.  The lifetime of polaritons $\gamma_C$ is controlled by the accuracy of the cavity DBRs and spans two orders of magnitude \cite{bec,nelsen}. The detuning between the cavity photon energy and the exciton resonance determines the proportion of photon and exciton in the polariton and, therefore,  the strength of the polariton-polariton and polariton-exciton interactions  and effective mass \cite{reviewCarusotto}.  The repulsive interactions between excitons and polaritons $g_R$ can be further tuned by the pumping geometry, for instance,  considering trapped condensates separated from the pumps \cite{baumbergTrapped}.

The building block of our network is a single stationary condensate described by a  wavefunction $\psi=\phi(r)$, created  by a spatially localised radially symmetric incoherent pumping source $P=p(r)$. For instance, a Gaussian pump $p(r)=A\exp[-w r^2]$ where $w$ determines the inverse width has been widely  used in experiments \cite{baumbergGeometrical,BerloffNatMat2017}. In what follows we assume that the pumping intensity $A$ is chosen so that $\phi$ is normalised, so that $\int_Q|\phi|^2\, d{\bf r}=1$, where $Q$ is the entire plane of the cavity.  We define  the corresponding stationary reservoir profile as the steady state of Eq.~(\ref{e2}), so that  $n_R(r)=n(r)=p/(b_0+b_1|\phi|^2)$.
 The networks of $N$ polariton condensates  are created at  lattice sites $i$  using a time and space varying pumping profile $P({\bf r},t)=\sum_{i=1}^N f_i(t)p(|{\bf r-r}_i|)$.   The total wavefunction $\psi$ and the reservoir density $n_R$ can be approximated by $\psi({\bf r},t)\approx \sum_{i=1}^N a_i(t) \phi(|{\bf r-r}_i|)$ and $n_R({\bf r},t)\approx \sum_{i=1}^Nk_i(t)n(|{\bf r-r}_i|)$ respectively. Such approximation is valid if the distance between the lattice sites exceeds the width of the condensate and reservoir \cite{GD-polariton}.  We use the shorth-hand notation $p_i\equiv p(|{\bf r-r}_i|)$ and similarly for $\phi$ and $n$.  We eliminate the spatial degrees of freedom by multiplying Eq.~(\ref{e1}) by $\phi_i^*$, Eq.~(\ref{e2}) by $|\phi_i|^2$ and integrating both equations over the plane of cavity $Q.$ \nb{Previously, we have used this approach on a single Ginzburg-Landau equation that is relevant to the description of  polariton condensates under several stringent assumptions: negligible blue-shift due to interactions of polaritons with the reservoir, short lifetime sample, fast reservoir relaxation, near threshold pumping intensity \cite{GD-polariton}.} Now we shall  drop such restrictions \nb{ and show how systems of different universality classes become relevant}. We use the smallness of the overlap integrals for the wavefunctions of the different lattice sites \cite{polaritonGraph} so  that $l_{ij}\equiv \int_Qn_i\phi_j\phi_i^*\, d{\bf r}\gg \int_Q\phi_j\phi_i^*\, d{\bf r},$ $l=l_{ii}\gg \int_Q n_i |\phi_j|^2\, d{\bf r},$ and $H=\int_Qn \nabla^2 \phi \phi^*\, d{\bf r} \gg \int_Qn_i \nabla^2 \phi_j  \phi_i^*\, d{\bf r}$ if $i\ne j.$ We also assume that for sufficiently smooth condensate profiles $\int_Q n_i \nabla^2\phi_j \phi_i^* \, d{\bf r} \ll l_{ij}.$ The dynamical equations on $\Psi_i(t)=a_i(t)\exp[-i d t]$ and $R_i(t)=l k_i $ become
 \begin{eqnarray}
 \dot{\Psi_i}&=& -i |\Psi_i|^2\Psi_i +hR_i\Psi_i+ (1-ig)[(R_i -1) \Psi_i + \sum_{j\ne i} J_{ij}\Psi_j],\label{ee1}\\
 \dot{R_i}  &=&b_0(\gamma_i- R_i -\xi R_i|\Psi_i|^2),\label{ee2}
 \end{eqnarray}  
 where we used the notation $d=\int_Q\phi^* \nabla^2 \phi\, d{\bf r},$  $h=\eta H/l,$ $\gamma_i=f_i \int_Qp|\phi|^2\, d{\bf r}/b_0, \xi=b_1 \int_Qn|\phi|^4\, d{\bf r}/l b_0,$ $J_{ij}=(R_i l_{ij} + R_j l_{ji}^*)/l.$ The energy relaxation parameter $\eta\ll 1$, therefore, $|H|<l$, so the term $|h| R_i \Psi_i $ will be neglected in comparison with $R_i\Psi_i$, whereas the imaginary part of $h$ will be assumed be absorbed by $g$. The coupling strength $J_{ij}$ is generally a complex number, so we write $J_{ij}\equiv {\cal J}_{ij} \exp[i v_{ij}]$  for real ${\cal J}_{ij}$ and $v_{ij}.$  In deriving Eqs.~(\ref{ee1}-\ref{ee2}) we neglected higher order nonlinearities in $\Psi_i$ in the view of their smallness close to the condensation threshold. 
 We consider several special cases of Eqs.~(\ref{ee1}-\ref{ee2}). 
 
 {\it Fast reservoir relaxation limit $b_0\gg 1$.} In this limit, the reservoir dynamics can be replaced with its steady state, so $R_i=\gamma_i/(1 + \xi|\Psi_i|^2) \approx \gamma_i - \xi \gamma_i |\Psi_i|^2$ reducing  the system of Eqs.~(\ref{ee1}-\ref{ee2})  to the single equation
 \begin{equation}
 \dot{\Psi_i}=i(g \xi \gamma_i-1)|\Psi_i|^2\Psi_i -\xi \gamma_i|\Psi_i|^2 \Psi_i  + (1 - ig) [(\gamma_i -1)\Psi_i + \sum_{j\ne i} J_{ij} \Psi_j].
 \label{sl}
 \end{equation}
 For uniform pumping $\gamma_i=\gamma$ this is a celebrated Stuart-Landau   system of coupled oscillators \cite{slref}.  This model can approximate a wide range of different oscillatory systems as it represents   the normal form of an Andronov-Hopf-bifurcation. Operating  close to an instability threshold lasers represent an example of the system  close to such a bifurcation.
 We substitute $\Psi_i(t)=\sqrt{\rho_i(t)}\exp[{\rm i} \theta_i(t)]$ into Eq.~(\ref{sl}) and separate real and imaginary parts to get 
\begin{eqnarray}
\frac{1}{2}\dot{\rho}_i(t)&=&(\gamma_i -1-\xi\gamma_i \rho_i) \rho_i + \sum_{j;j\ne i} \tilde{J_{ij}} {\sqrt{\rho_i\rho_j}}\cos(\theta_{ij}-v_{ij}+ \alpha),\label{rho}\\
\dot{\theta}_i(t)&=&(g \xi \gamma_i-1)\rho_i - g(\gamma_i - 1) -\sum_{j;j\ne i} \tilde{J_{ij}} {\frac{\sqrt{\rho_j}}{\sqrt{\rho_i}}} \sin(\theta_{ij}-v_{ij} + \alpha),\label{theta}
\end{eqnarray}
where $\theta_{ij}= \theta_i-\theta_j,$ $\tan\alpha=g$ and $\tilde{J_{ij}}={\cal J}_{ij}/\cos \alpha$. Note, that for the Gaussian pumping profile and wide reservoir, $|v_{ij}|\ll|{\cal J}_{ij}|$, so term $v_{ij}$  can be neglected since $l_{ij}\approx l_{ji}^*$ \cite{GD-polariton}. For other network geometries such assumption may not be valid in which case we can absorb $v_{ij}$ into $\alpha_{ij}=\alpha-v_{ij}$.
 
 Experimentally, the feedback can be applied to bring all the sites  to the same density $\rho_i(t)=|\Psi_i(t)|^2=\rho_{\rm th}$ by combining Eq.~(\ref{sl}) with an equation on the pumping adjustments
 \begin{equation}
 \gamma_i'(t)=\epsilon[\rho_{\rm th} - \rho_i(t)],
 \label{control}
 \end{equation}
 where the parameter $\epsilon$ characterises the rate of such adjustment or its discrete version applied at discrete times $t_n$ (more appropriate for the current experimental control techniques), so that $\gamma_i(t_n< t\le t_{n+1})=\gamma_i(t_n) + \epsilon (t_{n+1}-t_n) (\rho_{\rm th} - \rho_i(t_n)).$ Under this control, close to the threshold  $\rho_i\approx \rho_{\rm th}$ and Eqs.~(\ref{rho}-\ref{theta}) reduce to a single equation
 \begin{equation}
 \dot{\theta}_i(t)=(g \xi \gamma_i-1)\rho_{\rm th} - g(\gamma_i - 1) -\sum_{j;j\ne i} \tilde{J_{ij}}  \sin(\theta_{ij} + \alpha).\label{theta2}
\end{equation}
 This is the  Sakaguchi-Kuramoto model of coupled oscillators \cite{sakaguchi}  with $\alpha$ representing a phase lag. Synchronisation and desynchronisation in this system has been extensively studied in the contexts as vastly different as a network of  Wien-bridge oscillators in an experimental regime for which they can be approximated as phase oscillators \cite{wien},  power grids consisting of many oscillating generators \cite{powergrid}, and earthquake sequencing studies \cite{earth}. The phase lag appears as a result of synaptic organisations in neuroscience systems, time delays in sensor networks, or transfer conductances in power networks.  The  Sakaguchi-Kuramoto  model is a  special case of the Winfree model  with delta-function pulse shape $W_1(\theta)$ and a sinusoidal response
curve $W_2(\theta)$, so that  $\dot{\theta_i} =\omega_i + W_2(\theta_i) \sum_{i=1}^NW_1(\theta_j).$ If the coupling is sufficiently weak and the oscillators are nearly identical, the phase can be replaced by its average over an entire period of oscillations leading to the Sakaguchi-Kuramoto model.
 
 If $g=0$ ($\alpha=0$), then Eq.~(\ref{theta2}) reduces to the paradigmatic Kuramoto model \cite{kuramoto,kuramoto2} that was a first tractable mathematical model for describing how coherent behaviour emerges in complex systems. This model exhibits a phase transition at a critical
coupling, beyond which a collective behaviour is achieved. In our case, all natural frequencies are identical (and equal to $\rho_{\rm th}$) and the equation describes the negative gradient flow $\dot{\theta} = -\partial U(\theta)/\partial \theta$ for the smooth function $U(\theta)=-\sum_{i,j} \tilde{J_{ij}} \cos \theta_{ij}.$  Therefore,  by LaSalle Invariance Principle (e.g. in \cite{khalil}) every trajectory converges  to a minimum of the XY Hamiltonian  $H_{XY}=-\sum_{i=1}^N\sum_{j=1}^N \tilde{J_{ij}} \cos(\theta_i - \theta_j)$. In case of $\mathbb{S}^1$-synchronizing graphs  all critical points  are hyperbolic, so the synchronised state is the global minimum of $U(\theta)$, and all other critical points are local maxima or saddle points \cite{s1graphs}. For arbitrary graphs the global minimum can be achieved by implementing the lowest pumping intensity that leads to  threshold  $\rho_{\rm th}$ \cite{GD-polariton}.  Polaritons graphs as global minimisers of the XY Hamiltonian has been theoretically justified and experimentally realised in our previous work \cite{BerloffNatMat2017,polaritonGraph}.  Using a resonant pumping in addition to nonresonant one (adding the terms proportional to $\Psi_i^*$ to the right-hand side of Eq.~(\ref{e1})) allows one to minimise the Ising or Potts Hamiltonians \cite{GD-resonant}. Several  other  driving-dissipative platforms exploited this idea for  minimisation  of spin Hamiltonians:  injection-locked laser systems \cite{yamamoto11},
the networks of optical parametric oscillators, \cite{yamamoto14, yamamoto16a, yamamoto16b,takeda18},  coupled lasers \cite{coupledlaser},  and photon condensates \cite{KlaersNatPhotonics2017}. As the analogy with the coupled oscillators we presented above suggests, the minimisation of a spin Hamiltonian is realised  for the type of couplings that allow for the flow to be represented by the negative gradient flow with {\it real} function $U$. This condition is satisfied only if the coupling matrix $J_{ij}$ in Eq.~(\ref{sl}) is self-adjoint ${\cal J}_{ij}={\cal J}_{ji}$ and $v_{ij} = - v_{ji}$. For instance, if the couplings in Eq.~(\ref{sl}) are of  a pure Josephson type (e.g. $\dot{\Psi_i} =\cdot \cdot\cdot +{\rm i} \sum_j K_{ij} \Psi_j$  with real couplings $K_{ij}$) or have a non-negligible $g$,  such a network will not necessarily minimise  spin or any other Hamiltonian.

In addition, the parameter $g$ has a destabilising effect on the fixed points of Eq.~(\ref{theta2}). Different   $\gamma_i$ that have to be maintained to allow all densities to reach the same value $\rho_{\rm th},$ provide each lattice element with its own ``natural frequency," $\omega_i=(g \xi \gamma_i-1)\rho_{\rm th} - g(\gamma_i - 1),$ and, therefore, favour desynchronisation. In the network described by Eq.~(\ref{theta2})    synchronisation occurs when the coupling dominates the dissimilarity introduced by natural frequencies and the phase lag. The smaller is $g$ the more likely the global synchronisation is achieved. Concise  results for complex networks are known  for specific topologies such as, for instance, complete graphs, highly symmetric ring or linear graphs, acyclic graphs, and complete bipartite graphs  with uniform weights. 
Also, Sakaguchi phase lag parameter $\alpha$ contributes to desynchronisation as it provides attraction and repulsion between the oscillator phases similar to the  coupling time delay. The dependence of synchronisation and desynchronisation in polariton condensates on $g$ has been noted experimentally, but the reasons have not been previously identified \cite{wertz,manni,pendulum,ostrovskaya2018}. Such a behaviour, however, is easily explained from the point of view of the dynamics of coupled oscillators.

\begin{figure}[h!]
	\centering
\includegraphics[width=8.6cm]{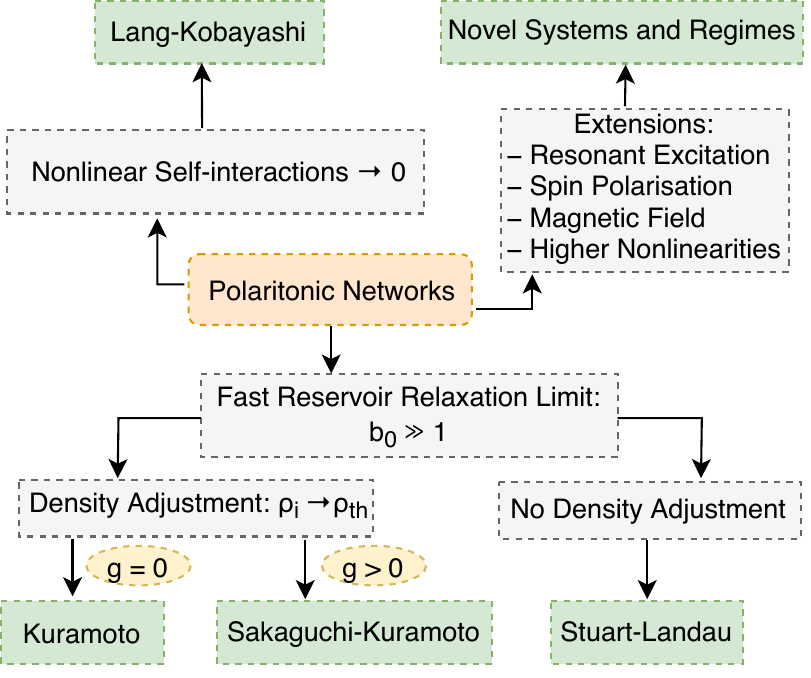}
	\caption{The polaritonic networks described by Eqs.~(\ref{ee1},\ref{ee2}) can lead to the Lang-Kobayashi model in the absence of the nonlinear self-interaction term or to Kuramoto/Sakaguchi-Kuramoto/Stuart-Landau models of coupled oscillators in the limit of fast reservoir relaxation. The new regimes are expected to appear due to strong polariton-polariton interactions or  once the experimental controls such as resonant excitation pump, spin polarisation, magnetic field, or combination of different sub-lattices are considered in polaritonic networks. }
    \label{Fig0_scheme}
\end{figure}

{\it Slow reservoir relaxation limit.} Eq. (\ref{sl})  describes the direct coupling scheme: pumping at the mean field, calculated algebraically from the states of all oscillators, enters the coupling. The coupling scheme of Eqs.~(\ref{ee1}-\ref{ee2}) is more complex: the mean field  acts on the reservoir densities that obey its own nonlinear differential equations, and the acting force is a function of the reservoir state. This is similar to the famous example of synchrony on London's Millennium Bridge  where  equations for the swinging mode of the bridge are coupled to the equations on individual pedestrians \cite{bridge}, or to electronic or electrochemical oscillators that are   coupled through the common macroscopic current or voltage, which obeys macroscopic equations describing the coupling circuit \cite{electric}. By tuning the photonic component of polaritons one can change the  polariton-polariton interactions up to 4 orders of magnitude \cite{ostrovskaya2018} which allows one to neglect the term $|\Psi_i|^2 \Psi_i$, so that Eqs.~(\ref{ee1}-\ref{ee2}) become  similar to the Lang-Kobayashi equations (with $\Psi_i$ replaced by the electric field and $R_i$ by the population inversion of the $i-$th laser) obtained using Lamb's semiclassical laser theory and capable of describing the dynamical behavior of coupled lasers \cite{lang80,acebron}. We summarise all the regimes and models described above schematically in Figure~\ref{Fig0_scheme}.

The flexibility  to tune the system parameters, the shape and geometry of the polariton lattice \cite{BerloffNatMat2017},  existence and tunability of  the nonlocal couplings beyond the next neighbour interactions \cite{exotic}, strong self-interactions of  polariton condensates allow one to not only recreate the intriguing patterns, states and structures that fascinated the nonlinear dynamics community in the last couple of decades but also  to enter novel  regimes.

To illustrate this  we consider two polaritonic networks configurations where the dynamic behaviour is the result of the interplay between two or more subsystems that belong to  different types of  coupled oscillators. 

\begin{figure}[b!]
	\centering
\includegraphics[width=10cm]{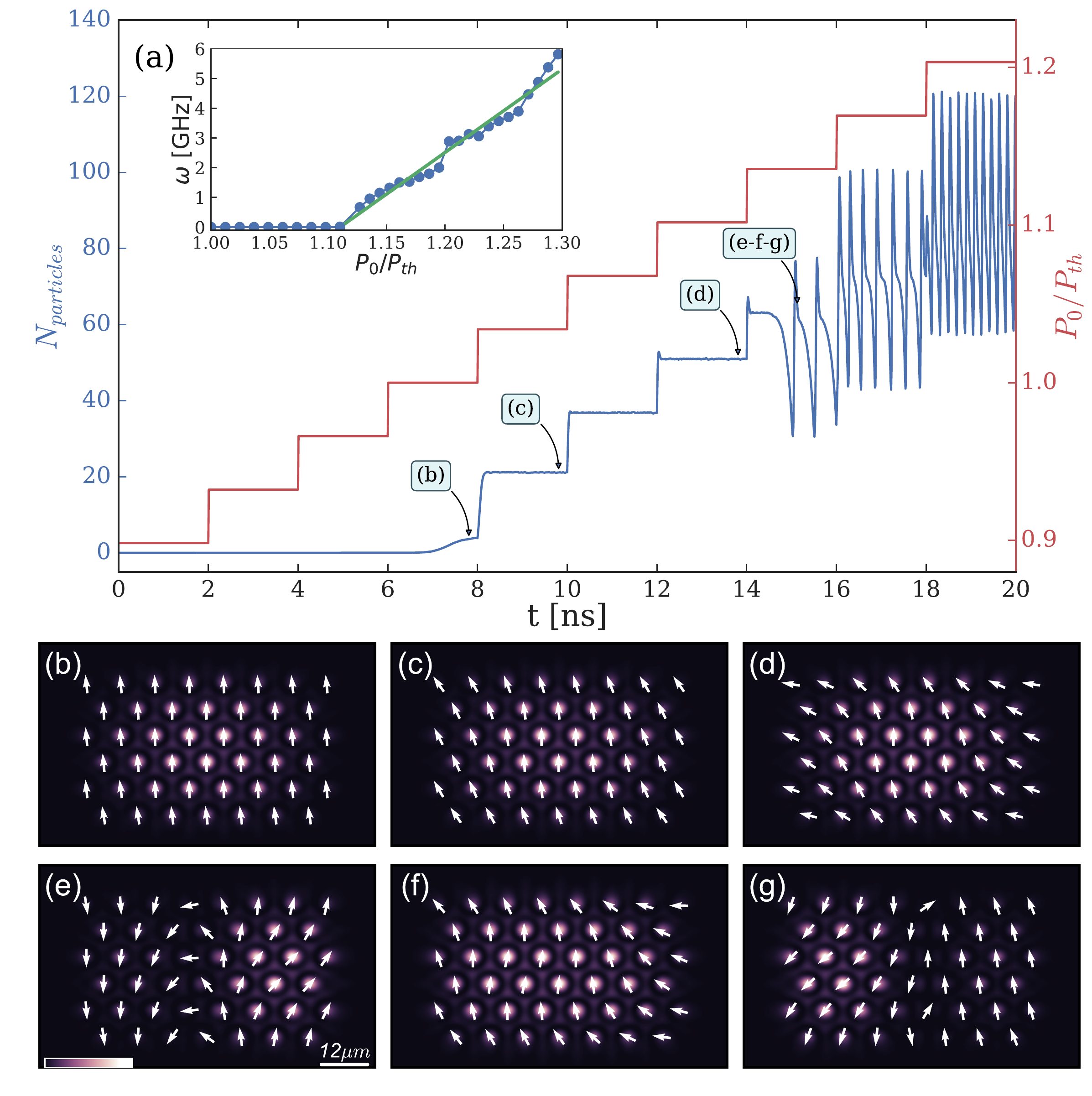}
	\caption{The results of numerical integration of Eq.~(\ref{e1} -\ref{e2}) in the fast reservoir relaxation limit for 45 polariton condensates arranged in the regular triangular lattice with a lattice constant $d = 9 \mu m$. (a) The total number of particles $N_{particles}$  as a function of time $t$ (blue line). The time-dependent pumping intensity increases above the threshold  $P_{0}/P_{\rm th}$ as shown in red, where $P_{\rm th} = 1.18$. The density distribution and phases of individual condensates are shown in (b-g). (b) The density and phases of the stationary lowest energy state at $P_{\rm th}$ with ferromagnetic couplings. (c-d) The spin configuration is changing for the higher pumping intensities as the DMIs become noticeable. (e-f-g) Time snapshots over the period of density oscillations at $P_0 = 1.14 P_{\rm th}$  demonstrate the collective density oscillations with a temporary formation of the domain wall. The inset shows the oscillation frequency of the number of particles $N_{particles}$  as a function of $P_0/P_{\rm th}$. The transition point between the stationary  state and the collective periodic density oscillations is at $P_0 = 1.11 P_{\rm th}$. The frequency of oscillations increases linearly for $P_0>1.11 P_{\rm th}$. For  $P_0>1.3 P_{\rm th}$ the system exhibits chaotic density oscillations. Random initial conditions and white noise were used in numerical simulations with $g= 0.1$, $b_0=0.3$, $b_1=0.009$, $\eta=0.12$, $w=1.33$, $P({\bf r})=\sum_{i=1}^N P_0 \exp[-w ({\bf r- r}_i)^2]$.}
    \label{Fig1}
\end{figure}

{\it Triangular lattice in fast reservoir relaxation regime, low $g$, short lifetime, constant pumping.} Any finite two-dimensional polariton lattice with the same pumping intensity  across all the sites gives rise to an example of an interplay between Kuramoto (for $g=0$) or Sakaguchi-Kuramoto (for $g\ne 0$) oscillators away from the network boundaries with the Stuart-Landau oscillators at the sites close to the boundary. When all the lattice sites are equally pumped, the  internal sites  have the same number of neighbours and,  therefore, receive equal particle fluxes from them. On the boundary, however, the number of neighbours is diminished, so fewer fluxes bring fewer particles decreasing the density of the oscillators. The difference in the ``natural frequency"  (given by the first term of the right-hand side of Eq.~(\ref{theta})) between the boundary and bulk condensates tends to desynchronise the lattice, but for lower pumping intensities this effect is overwhelmed by the synchronisation effect of the couplings, so  the lattice is frequency synchronised (see Fig.~\ref{Fig1}(b)). As the pumping intensity is increased further above the threshold, the frequencies may  still be synchronised, but the phases corresponding to a spin glass emerge as shown in Fig.~\ref{Fig1}(c,d). The appearance of the spin glass is due to nonlocal coupling and an increasing role of Dzyaloshinskii-Moriya interactions (DMIs), since the density misbalance between the sites makes the interaction directional and, therefore, of the DMIs type \cite{ourDM, SkyrmionsNatureNano2013}. For even higher pumping intensities the desynchronisation takes place (Fig.~\ref{Fig1}(e-g)) resulting in a peculiar quasi-one-dimensional form of large-scale collective density oscillations in the direction of the larger axis of symmetry of the lattice. The collective density oscillates between left and right parts of the lattice with the appearance (and disappearance)  of $\pi-$ phase difference between two clusters, indicating temporal formation of the domain wall. Figure \ref{Fig1} illustrates these regimes using the full spatially resolved simulations of the  polaritonic networks (Eqs.~(\ref{e1},\ref{e2})) in the  fast reservoir relaxation limit. These states  are  robust, sustain an external noise, exist for a wide range of parameters, and   reached from random initial conditions. The collective density oscillations also appear  when larger lattices are considered. 

We note that the system behaviour as the pumping intensity increases is reminiscent  of  the Mott metal-insulator transition. The frequency of the collective oscillations  quench  towards the  transition point (see the inset in Fig. \ref{Fig1}(a)) after which the oscillation frequency increases as a linear function, repeating the linear trend of the original Mott transition.  We can induce reversible changes between these two states of the system by increasing or decreasing pumping intensity. Such oscillations suggest an interesting application towards  implementation of polariton transistors \cite{polaritonTransistor} that operate  in the gigahertz range.

\begin{figure}[b!]
	\centering
\includegraphics[width=8.6cm]{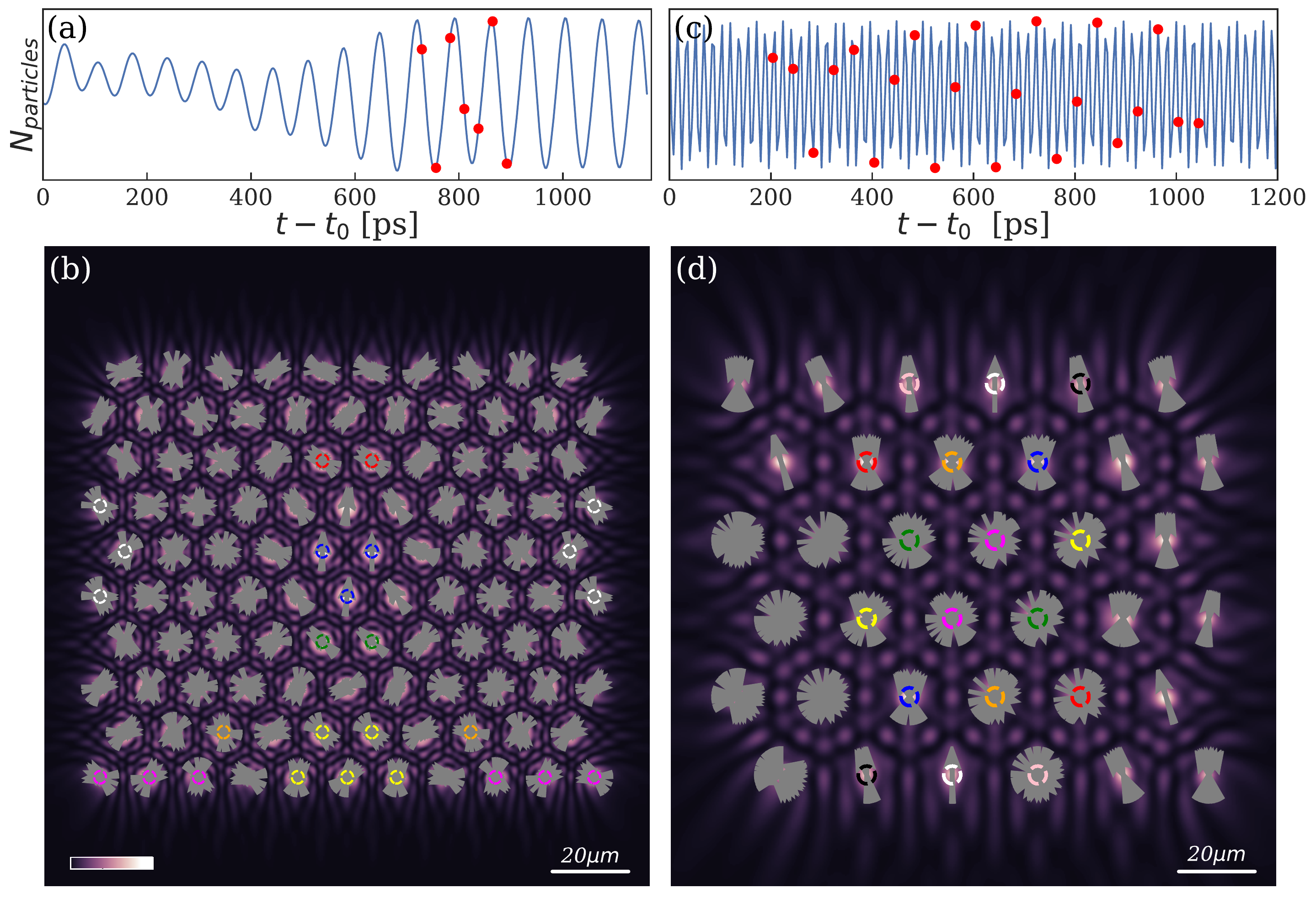}
	\caption{(a,c) The numbers of particles $N_{particles}$ as functions of time $t-t_0$ calculated by numerical integration of  Eq.~(\ref{e1} -\ref{e2}) in (a) fast  [(c) slow] reservoir relaxation limits for (a) 105 [(c) 36] polariton condensates arranged in the regular triangular lattice with a lattice constant (a) $d = 13 \mu m$ [(c) $d = 22.5 \mu m$].  (b,d) The time-averaged density and instantaneous relative phases of individual condensates. Relative phases  are plotted for times depicted with red dots in (a,c), respectively.  The clusters of synchronised condensates are indicated by dashed circles of the same colour. The numerical parameters in (a,b) are the same as in Fig.~\ref{Fig1} with $P_0 = 9.5$, the parameters in (c,d) $g= 0.1$, $b_0=0.05$, $b_1=200$, $\eta=0.01$, $w=0.8$, $P({\bf r})=\sum_{i=1}^N P_0 \exp[-w ({\bf r- r}_i)^2]$, $P_0 = 100$; $t_0$ is about $200ps$ in both cases.}
    \label{FigChimera}
\end{figure}

{\it Triangular lattice, in both fast and slow reservoir relaxation regimes, low $g$, short and long lifetime, constant pumping.} Our second example concerns cluster synchronization --  a particular synchronization phenomenon that requires that synchronization occurs in each group, but there is no synchronization among the different groups \cite{cluster}.  The importance of cluster synchronization has been found in various applications including biological sciences  and communication engineering, and   various control schemes were designed  to drive the network to cluster synchronization \cite{cluster2}. In polaritonic networks, cluster synchronization presents itself even in the case of equally pumped lattices. To illustrate cluster formation, we consider regular triangular lattices and numerically integrate  Eqs.~(\ref{e1},\ref{e2}) in two opposite limits: fast (Fig.~\ref{FigChimera}(a)) and slow (Fig.~\ref{FigChimera}(b)) reservoir relaxation times. During a fraction of a nanosecond the system is frequency locked into several clusters, some of them are shown on Fig.~\ref{FigChimera}(b,d) with dashed circles of the same color. The cluster state on Fig.~\ref{FigChimera}(a,b)  is a long-lived transient state  that after a few nanoseconds evolves into chaotic oscillations. However, both reported states are stable to addition of the random noise and to intrinsic roughness of the sample modelled by adding noise potential to the right hand side of Eq.~(\ref{e1}) and, therefore, may be detected experimentally.

The established links between polaritonic networks and other coupled oscillators systems suggest that these two examples are infinitely far from describing all possible dynamical  regimes and  hint that other intriguing states can be obtained. A particular interest is in finding Chimera states.  A Chimera state is a spatio-temporal pattern in a network of identical coupled oscillators in which some of the oscillators  synchronise (become frequency locked) while others remain incoherent (desynchronised)  \cite{chimera1,stationarypattern}. The  simplest model that supports such  states was reported as a pair of Sakaguchi-Kuramoto oscillator populations in which each oscillator has different coupling to others in the same group then to those in another group \cite{chimeraStrogatz}. The type of the pattern depends strongly on the parameter $g$, and, therefore, the phase lag $\alpha$. The spiral pattern of chimeras appears when $\alpha$ is close to zero whereas the spot chimeras only appear when $\alpha$ is close to $\pi/2$ \cite{chimerasReview}. The `turbulent chimeras'  were observed in some special cases of nonlocally coupled Stuart-Landau oscillators in which regions of local synchronisation appeared and vanished  randomly over time \cite{turbulentChimeras}. The  Stuart-Landau system Eq.~(\ref{sl}) has also been shown to produce a wealth of various stationary and dynamical behaviours, e.g.  amplitude death, Hopf oscillations, large oscillations, quasi-periodicity, and chaos \cite{matthew91}. Polaritonic networks  is a promising platform to study the wealth of these phenomena.

To summarize,  we propose polaritonic networks as  a paradigm for the dynamics of disparate systems of coupled oscillators. \nb{The dynamics of coupled oscillators often appear in the context of lasers and other driven-dissipative systems. It is therefore not surprising that systems of coupled oscillators are closely related to polaritonic systems.} However, we show that depending on the system parameters and experimental controls used such networks can not only reproduce the behaviour of various known coupled oscillators and allow one to address  intermediate regimes between different types of systems, but also to  dramatically increase the physics of the system by combining different types of oscillators together in one interacting platform. This opens intriguing possibilities for entering novel hybrid regimes that have never been implemented before. Polaritonic network is a flexible and robust platform for achieving this.


\end{document}